# Advanced Security Threat Modelling for Blockchain-Based FinTech Applications

**Serhan W. Bahar**

*London, United Kingdom*

**Abstract** Cybersecurity threats and vulnerabilities continue to grow in number and complexity, presenting an increasing challenge for organizations worldwide. Organizations use threat modelling and bug bounty programs to address these threats, which often operate independently. In this paper, we propose a Metric-Based Feedback Methodology (MBFM) that integrates bug bounty programs with threat modelling to improve the overall security posture of an organization. By analyzing and categorizing vulnerability data, the methodology enables identifying root causes and refining threat models to prioritize security efforts more effectively. The paper outlines the proposed methodology and its assumptions and provides a foundation for future research to develop the methodology into a versatile framework. Further research should focus on automating the process, integrating additional security testing approaches, and leveraging machine learning algorithms for vulnerability prediction and team-specific recommendations.

## 1. Introduction

The financial impact of cybersecurity incidents is expected to reach a staggering $6 trillion in 2021 [1]. Cybersecurity Ventures Special Report also anticipates that the cost of cybercrimes will escalate to $10.5 trillion by 2025 [2]. This rapid surge in hacking activities poses significant threats to individuals, corporations, and governments alike. Among the affected sectors, the blockchain industry has been particularly vulnerable, with hackers stealing over $4 billion worth of cryptocurrencies in 2021 alone [3]. These statistics account only for direct monetary losses. The overall cost is substantially higher when factoring in the broader business impact of these cybercrimes.

Despite the advantages of permissionless and decentralized public blockchains, one notable characteristic is the non-reversible nature of transactions. Once a transaction is digitally signed and added to a block, reversing it becomes virtually impossible. Consequently, stolen funds are irretrievable unless the hacker chooses to return them.

In light of these challenges, cybersecurity professionals must prioritize metrics within the blockchain industry. Regrettably, the complex landscape of cybersecurity often resembles navigating through a dark forest without appropriate metrics and feedback mechanisms, akin to flying blind.

The importance of metrics and feedback mechanisms is also emphasized in the paper "Science of Security Hard Problems," specifically in the category of "Security-Metrics-Driven Evaluation, Design, Development, and Deployment" [4]. The objective is evident from the report's definition of the hard problem: "Develop security metrics and models capable of predicting whether or confirming that a given cyber system preserves a given set of security properties (deterministically or probabilistically), in a given context" [4].

"When you can measure what you are speaking about, and express it in numbers, you know something about it; but when you cannot measure it, when you cannot express it in numbers, your knowledge is of a meagre and unsatisfactory kind: it may be the beginning of knowledge, but you have scarcely, in your thoughts, advanced to the stage of science, whatever the matter may be" [5]

The significance of metrics has inspired us to present this paper, proposing a metric-based feedback method for threat modelling that utilizes metrics from security assessment and testing results, such as bug bounties.

The remainder of this paper is organized as follows: In Section 2, we review the literature on one of the most prevalent threat modelling methodologies, STRIDE. Additionally, we examine bug bounty methodologies for blockchain applications. We also explore the current progress of blockchain application security industry standards in the literature, such as the SWC Registry and

.

I

SCSVS. In Section 3, we propose our metric-based feedback methodology. Finally, in Section 4, we present the paper's conclusions along with potential use cases and directions for future work.

## 2. LITERATURE REVIEW

### 2.1. Stride Threat Modeling

One of the most widely recognized and mature threat modelling methodologies is STRIDE, developed by Loren Kohnfelder and Praerit Garg. Since its introduction in 1999, STRIDE has been a prominent fixture in the security industry, and Microsoft incorporated it into their corporate structure in 2002 [6]. Various implementations of STRIDE exist, but the core principle remains consistent: identify risks before others do. STRIDE is an acronym for Spoofing, Tampering, Repudiation, Information Disclosure, Denial of Service, and Elevation of Privilege. (Refer to Table-1 for details)

TABLE 1
STRIDE THREAT MODEL ELEMENTS

| Property | Threat | Definition |
|---|---|---|
| Authentication | **S**poofing | Impersonating something or someone else. |
| Integrity | **T**ampering | Modifying data or code. |
| Non-repudiation | **R**epudiation | Claiming to have not performed an action. |
| Confidentiality | **I**nformation Disclosure | Exposing information to someone not authorized to see it. |
| Availability | **D**enial of Service | Deny or degrade service to users. |
| Authorization | **E**levation of Privilege | Gain capabilities without proper authorization. |

*Note.* Adapted from STRIDE Chart - Microsoft Security Blog, by Microsoft, 2007 (https://www.microsoft.com/en-us/security/blog/2007/09/11/stride-chart/)

STRIDE relies on creating Data Flow Diagrams (DFDs) to visualize the high-level architecture and define system assets and their relationships.

There are four steps to implementing STRIDE. In step 1, a DFD is created to define the scope and establish relationships between the assets. In step 2, DFD elements are mapped to the threat categories presented in Table-1. Step 3 involves identifying all the threats. The fourth and final step entails documenting the threats and describing security requirements or controls [7].

One of the most important research about STRIDE shows that it has mainly two issues. The first one is that the number of threats depends on the system's complexity. If the system's complexity increases, the number of threats will overgrow. The second one is that the STRIDE method has a moderately low rate of false positives and a moderately high rate of false negatives [7].

Threat modelling is typically situated in the architecture and design phase of the software development lifecycle [8].

### 2.2. Bug Bounty Methodologies

Identifying bugs and vulnerabilities is generally situated in the testing phase of the software development lifecycle [8]. Various methodologies and approaches exist for testing and discovering vulnerabilities in software, with bug bounty programs being an increasingly popular method [9].

Bug bounty programs provide a structured approach to attract white hat hackers by incentivizing them to identify bugs and vulnerabilities in software. These incentives can include monetary rewards, merchandise, or recognition in a hall of fame.

Bug bounty programs can be divided into three categories. The first type, Institutional BBPs, includes programs like Google's and Microsoft's Bug Bounty Programs. These programs are hosted directly by the companies that own the software. The second type, Platform BBPs, involves companies such as HackerOne, Bugcrowd, HackenProof, and Immunify. These companies act as intermediaries, connecting hackers and businesses and fostering a trust model between them. Lastly, Private Intermediary BBPs, like Zerodium and Sonosoft, aim to provide higher rewards while working with more tailored solutions [10].

Additionally, four variants of bug bounty programs can be observed in Table-2 below.

TABLE 2
BUG BOUNTY VARIANTS

| Type of BBP | Description | Phase |
|---|---|---|
| Invite only | Managed either by firm or platform wherein only top contributors are invited. | Beta launch or after product launch |
| Fuzzing competition | For internal testing. Firms must provide access to internal sources. | Prebeta or postlaunch |
| Open-ended BBP | Ongoing BBPs after product launch | Postlaunch |
| Short-time-frame competitions | Special instances, such as product integratin etc. prompt a short-term event | One-time competitions, postlaunch etc. |

*Note.* Adapted from (Bug Bounty Programs for Cybersecurity: Practices, Issues, and Recommendations) by Malladi S, Subramanian H, 2020

According to research, bug bounty programs generate an average of 0.429 new valid report submissions per day and 156 valid reports per year. Furthermore, with an average of 156 reports per year, an organization can expect to uncover 13 critical vulnerabilities [9].

Running a bug bounty program is an effective way to identify and address vulnerabilities. However, without determining the root cause of these vulnerabilities, it is possible to fix the current issue, but there is no guarantee that similar vulnerabilities will not resurface. Unfortunately, current bug bounty methodologies do not offer direct feedback to threat models for pinpointing the root causes of vulnerabilities. As a result, the lack of feedback mechanisms remains a limitation of bug bounty programs.

2.3. SWC Registry

The SWC Registry (The Smart Contract Weakness Classification Registry) is an adaptation of the weakness classification scheme, tailored explicitly for smart contracts. It employs the terminology of the CWE (Common Weakness Enumeration) while highlighting variants unique to smart contracts [11].

The project's main objective is to establish a method for classifying vulnerabilities in smart contracts. Additionally, it aims to develop a common language for articulating security concerns in the architecture, design, or code of smart contract systems [11].

The SWC Registry organizes vulnerabilities into four categories: ID, Title, Relationships, and Test Cases. An example of this classification can be seen in Table-3 below.

TABLE 3
SMART CONTRACT WEAKNESS CLASSIFICATION AND TEST CASES

| ID | Title | Relationships | Test cases |
| --- | --- | --- | --- |
| SWC-136 | Unencrypted private data on-chain | CWE-767: Access to Critical Private Variable via Public Method | odd_even.sol |
| SWC-135 | Code with no effects | CWE-1164: Irrelevant Code | wallet.sol |

*Note.* Adapted from [14] (https://swcregistry.io/)

The SWC Registry serves as an excellent example of vulnerability classification for blockchain applications.

2.4. SCSVS

The SCSVS (Smart Contract Security Verification Standard) is a checklist designed to standardize security for smart contracts. In addition, it aims to help mitigate known vulnerabilities, covering aspects from design to implementation [12].

Similar to OWASP v4.0, the objectives of SCSVS include:
- Help to develop high-quality code of the smart contracts.
- Help to mitigate known vulnerabilities by design.
- Provide a checklist for security reviewers.
- Provide a clear and reliable assessment of how secure a smart contract is with the percentage of SCSVS coverage [12].

SCSVS advised of usage of their checklist as:

"As a starting point for formal threat modelling exercise. As a measure of your smart contract security and maturity. As a scoping document for a smart contract penetration test or security audit. As a formal security requirement list for developers or third parties developing the smart contract for you. As a self-check for developers. To point out areas which need further development in regards to security" [12].

There are 14 sub-sections of the list:

- V1: Architecture, Design and Threat Modelling
- V2: Access Control
- V3: Blockchain Data
- V4: Communications
- V5: Arithmetic
- V6: Malicious Input Handling
- V7: Gas Usage & Limitations
- V8: Business Logic
- V9: Denial of Service
- V10: Token
- V11: Code Clarity
- V12: Test Coverage
- V13: Known Attacks
- V14: Decentralized Finance

SCSVS is a good example of security checklists in blockchain applications.

3. METRIC-BASED FEEDBACK METHODOLOGY (MBFM)

3.1. Methodology Introduction

Numerous organizations employ bug bounty programs to address vulnerabilities. The bug bounty process is typically situated in the penultimate stage of the Secure Software Development Lifecycle (the testing phase), while threat modelling occurs in the second stage (the design phase) [8].

Integrating bug bounty submissions into regular threat model sessions has the potential to revolutionize the way that security teams approach vulnerability management. The traditional approach to vulnerability management has been focused on identifying vulnerabilities and patching them as they are discovered. However, this approach is reactive and does not address the underlying causes of vulnerabilities.

By incorporating bug bounty submissions into regular threat model sessions, security teams can better understand the root causes of vulnerabilities. This can allow them to take a more proactive approach to vulnerability management by identifying and addressing the underlying issues that are leading to vulnerabilities in the first place.

At the beginning of this section, we will present a sample threat model diagram and the proposed metric-based feedback methodology. Subsequently, we will outline our example assumptions towards the end of the section.

In our threat model diagram, we utilized the following drawing elements. (Refer to Figure-1 for details)

Figure 1
*Threat model diagram elements*

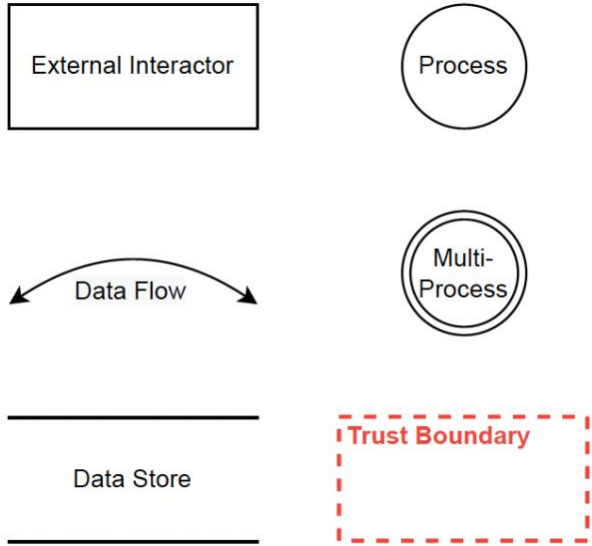

STRIDE threat models typically consist of three categories for labelling elements in the data flow diagram: Assets, Security Controls, and Threat Actors [7]. In the asset section, all essential application components, security controls, and threat actors will be listed and visualized. (Refer to Figure-2 for details)

Figure 2
*Threat model example*

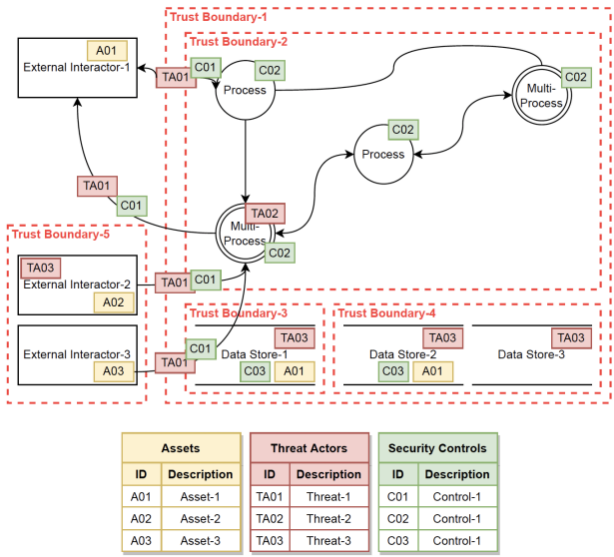

When products are available for bug bounties and receive valuable submissions, it becomes possible to analyze vulnerabilities and tag them with SWC Registry and SCSVS, shedding light on the initial root causes of these vulnerabilities. If this process is repeated quarterly or semi-annually, it can provide more insight into the root causes. However, identifying the root cause is not the sole benefit; it also allows for evaluating the accuracy of threat modelling, threat actors, and security controls. (Refer to Figure-3 for details)

Figure 3
*Threat model example with metric-based feedback methodology*

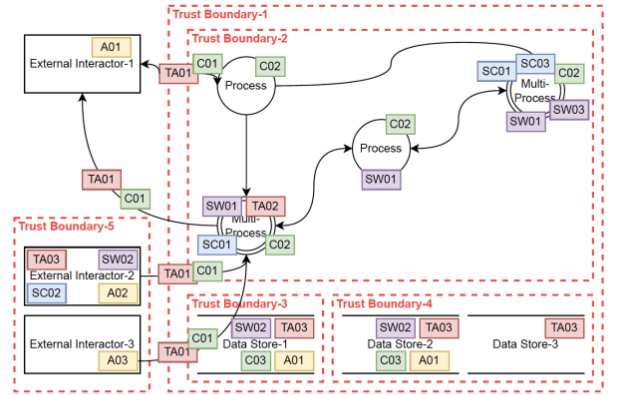

To implement a metric-based feedback methodology, the results of valid bug bounty submissions should be carefully reviewed and tagged with relevant categories from the SWC Registry and SCSVS. Subsequently, the findings' labels will be shared with threat modelling. In this example, it is possible to visualize the vulnerabilities within the diagram.

3.2. Assumptions

3.2.1. Threat models inherently involve numerous assumptions about threat actors and their actions. Establishing a feedback loop between bug bounties and threat models presents an opportunity to evaluate these assumptions using metrics. We presume that this methodology will reveal the accuracy of threat model outputs and potentially uncover overlooked assets in the diagram.

3.2.2. The Metric-Based Feedback Methodology (MBFM) could help identify which assets have the most vulnerabilities. By considering the severity and frequency of these vulnerabilities, security teams can prioritize tasks to enhance security controls.

3.2.3. Incorporating bug bounty results into threat modelling and reviewing the most common vulnerability categories from SWC Registry and SCSVS may provide valuable insights into the root causes. Consequently, management can make informed decisions about resource allocation based on these results.

3.2.4. Analyzing root causes using MBFM can yield insights into the performance of teams, such as developers or security teams. For instance, human resources departments could recruit employees with specific skills to address certain vulnerabilities or provide internal or external training to enhance the team's knowledge and ability to mitigate current vulnerabilities.

3.2.5. Unknown chronic issues might become evident by examining vulnerability frequencies. Monitoring the recurrence of specific vulnerabilities over time can help organizations identify persistent weaknesses that require attention and remediation efforts. This analysis can guide security teams in addressing these chronic issues more effectively and preventing their recurrence in the future.

## 4. Conclusion

The methodology presented in this paper aims to establish a foundation for future research and the development of a comprehensive framework for corporations and governments. However, further research and scientific experimentation are necessary to refine this methodology into a versatile framework.

### 4.1. Future Works

4.1.1. As the analysis and processes involved in this methodology require manual labour, some of these processes could be automated using software, reducing human effort and increasing efficiency.

4.1.2. If organizations adopting the MBFM anonymize and publicly share their results, it may be possible to create a comprehensive database of attack vectors and surfaces. This data could pave the way for future research in machine learning and big data science.

4.1.3. Analyzing teams' behaviours and strengths/weaknesses, along with the root causes of vulnerabilities, may enable the development of predictive mechanisms for future project planning through academic research.

4.1.4. In addition to bug bounty results, incorporating findings from penetration testing, security audits, and static/dynamic test analysis can provide a more holistic approach and improve the accuracy of the methodology.

4.1.5. Once sufficient high-quality data has been collected, machine learning algorithms can be employed to analyze the information. This analysis could facilitate the prediction of future vulnerabilities and the generation of team or software-specific recommendations.

4.1.6. Assumptions made within the methodology should be rigorously tested, and their benefits measured. Experimental designs and tests can be conducted with experimental groups, and the findings can be published in academic literature to contribute to the body of knowledge in this field.

In conclusion, this paper has proposed a metric-based feedback methodology (MBFM) to enhance threat modelling in the blockchain industry. The primary goal of this methodology is to improve the security posture of blockchain applications by incorporating the findings from bug bounty programs, security assessments, and testing results. Furthermore, by identifying the root causes of vulnerabilities and updating threat models with real-world data, organizations can develop more robust security controls and mitigate potential threats more effectively.

We have discussed the significance of existing threat modelling methodologies such as STRIDE and the role of bug bounty programs in identifying vulnerabilities in blockchain applications. Additionally, we have highlighted the importance of industry standards like the SWC Registry and SCSVS in classifying and mitigating vulnerabilities. By integrating these components into the MBFM, organizations can better understand their security landscape and make informed decisions about resource allocation, risk management, and future development.

Throughout the paper, we have presented several assumptions and potential benefits of the proposed methodology. However, it is crucial to validate these assumptions and measure the effectiveness of the MBFM through rigorous testing and experimentation. Future work in this area should focus on automating some of the processes involved, incorporating additional sources of security data, leveraging machine learning algorithms, and conducting academic research on team dynamics and root cause analysis.

Ultimately, the metric-based feedback methodology could revolutionise how organisations approach threat modelling and security management in the blockchain industry. By continuously updating threat models with real-world data and insights, security professionals can stay ahead of the ever-evolving threat landscape and ensure blockchain applications' integrity, confidentiality, and availability. Furthermore, as the industry matures and more organisations adopt blockchain technology, adopting such a methodology will become increasingly critical in maintaining the trust and fostering the continued growth of this transformative technology.

## Acknowledgements

The author extends his gratitude to Dr Yogachandran Rahulamathavan for his support throughout the development of this paper. We also acknowledge the support of the Institute for Digital Technologies at Loughborough University London.